\documentclass[baaa]{baaa}

\usepackage[pdftex]{hyperref}
\usepackage{subfigure}
\usepackage{natbib}
\usepackage{helvet,soul}
\usepackage[font=small]{caption}

\begin{document}


\journalvol{61A}
\journalyear{2019}
\journaleditors{R. Gamen, N. Padilla, C. Parisi, F. Iglesias \& M. Sgr\'o}


\contriblanguage{1}


\contribtype{2}

\thematicarea{7}

\title{Accretion models for LLAGNs: Model Parameter Estimation}
\subtitle{and Prediction of their Detectibility}


\titlerunning{Accretion models for LLAGNs}


\author{Bidisha Bandyopadhyay\inst{1}, Dominik R. G. Schleicher\inst{1}, Neil Nagar\inst{1}, Fu-Guo Xie\inst{2} \& Venkatessh Ramakrishnan\inst{1}}
\authorrunning{Bandyopadhyay et al.}


\contact{bidisharia@gmail.com}

\institute{
Departamento de Astronom\'ia, Facultad Ciencias F\'isicas y Matem\'aticas, Universidad de Concepci\'on, Av. Esteban Iturra s/n Barrio Universitario, Casilla 160-C, Concepci\'on, Chile \and
Key Laboratory for Research in Galaxies and Cosmology, Shanghai Astronomical Observatory, Chinese Academy of Sciences, 80 Nandan Road,
Shanghai 200030, China
}


\resumen{
El Event Horizon Telescope (EHT) brinda una oportunidad única para explorar la física de los agujeros negros supermasivos a través de la Interferometría de Línea de Base Muy Grande (VLBI), como la existencia del horizonte de eventos, los procesos de acreción y la formación de chorros. Construimos un modelo teórico que incluye un 'Advection Dominated Accretion Flow' (ADAF) y un modelo simple de un chorro que en al radio. Comparando la distribución de energía espectral (SED) predicha con los datos observados se obtiene las mejores estimaciónesde los parametros del modelo. Además, al investigar los perfiles de emisión radial de las predicciones del modelo en diferentes bandas de frecuencia, podremos predecir si el EHT puede resolverlos.}

\abstract{
The Event Horizon Telescope (EHT) provides a unique opportunity to probe the physics of supermassive black holes through Very Large Baseline Interferometry (VLBI), such as the existence of the event horizon, the accretion processes as well as jet formation. We build a theoretical model which includes an Advection Dominated Accretion Flow (ADAF) and a simple radio jet outflow. The predicted spectral energy distribution (SED) of this model is compared to observations to get the best estimates of the model parameters.
Also the model-predicted radial emission profiles at different frequency bands can be used to predict whether the inflow can be resolved by the EHT.   
}



\maketitle

\section{Introduction}
\label{S_intro}
Active Galactic Nuclei (AGN) are among the brightest sources in the sky. It is well established that the accretion processes around compact objects are the most energetic processes in the universe, with an efficiency higher than even nuclear fusion. Their central engines are therefore expected to be the primary powering source illuminating the AGN. The existence of a highly massive compact object in the center of our galaxy (Sgr A*) has been confirmed from the observations of stellar motions near the centre \citep{2012Sci...338...84M}. Based on the predictions of General Relativity, this compact object is usually assumed to be a black hole, i.e. an object with an event horizon, from which not even light can escape. To rule out the existence of a surface for the compact object in Sgr A* and in the Virgo cluster CD galaxy M87, \citet{Broderick09, Broderick15} pursued a comparison of the observed fluxes in the central regions and the expected fluxes in the presence of a putative surface. They have shown that, for realistic accretion rates, the existence of such a surface is highly implausible. However, this still presents an indirect argument based on assumptions. The detection of the shadow of the nuclear black hole would provide direct and firm evidence of the existence of a horizon. Imaging the shadow of the supermassive black holes in Sgr A$^*$ and M87, thus detecting their horizons, is thus one of the main goals of the Event Horizon Telescope (EHT) \footnote[5]{Webpage EHT: http://www.eventhorizontelescope.org/} which has a resolution of 15-20 micro-arcseconds. In addition, the EHT will study the accretion of supermassive black holes in other nearby Low-Luminosity AGN (LLAGN). While the Global 3-mm VLBI Array (GMVA) \footnote[6]{Webpage GMVA: https://www3.mpifr-bonn.mpg.de/div/vlbi/globalmm/} and the European VLBI Network (EVN) \footnote[7]{Webpage EVN: http://www.evlbi.org/} observe at lower frequencies and offer lower resolutions (50-70 micro-arcseconds), observing nearby LLAGN with them will enable the characterization of emission from a greater part of the accretion disk.
                       
\begin{figure*}
\begin{center}
{ \includegraphics[height=2.2in,width=2.2in]{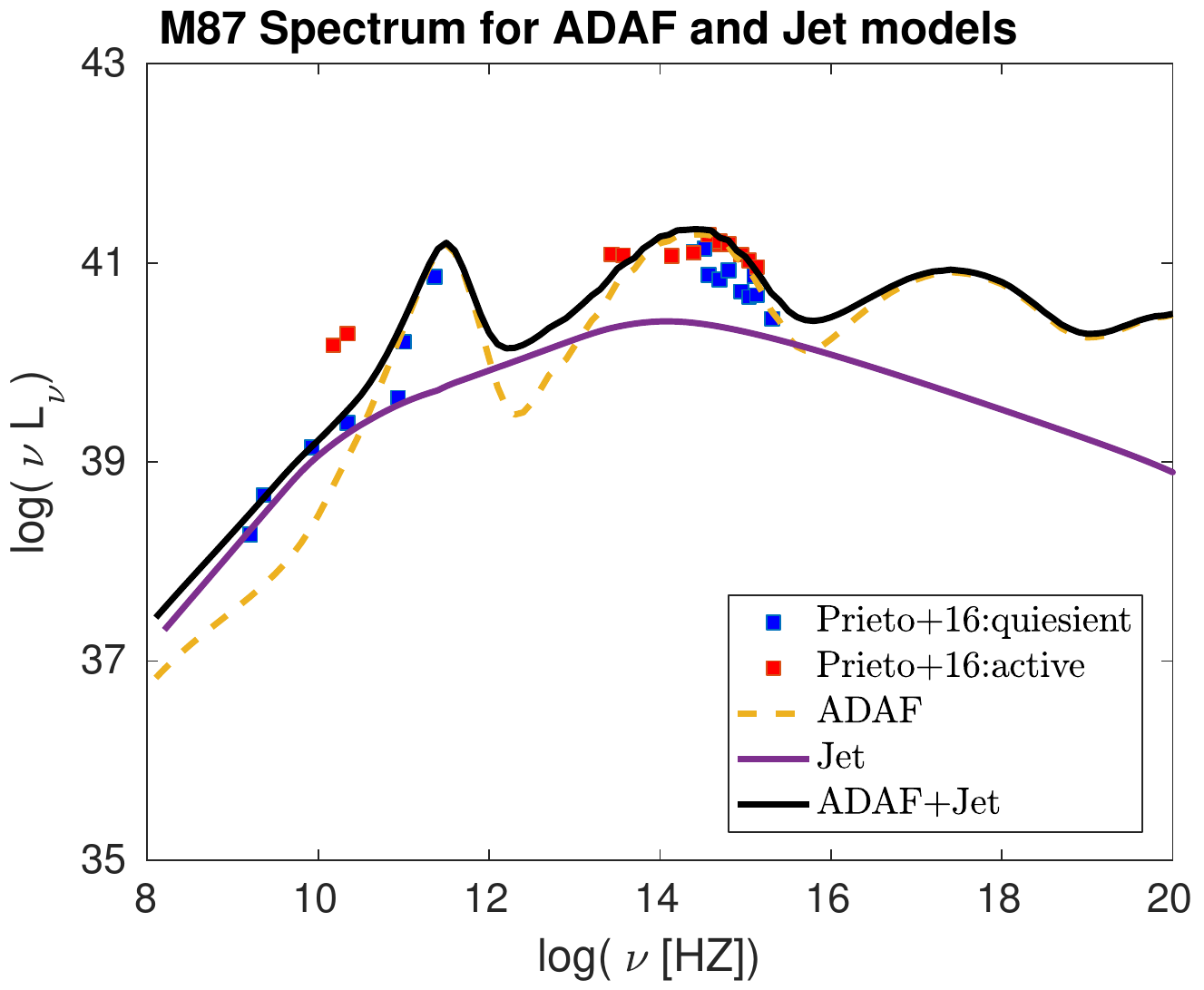}}~~~~~~~~~~~~
{ \includegraphics[height=2.2in,width=2.2in]{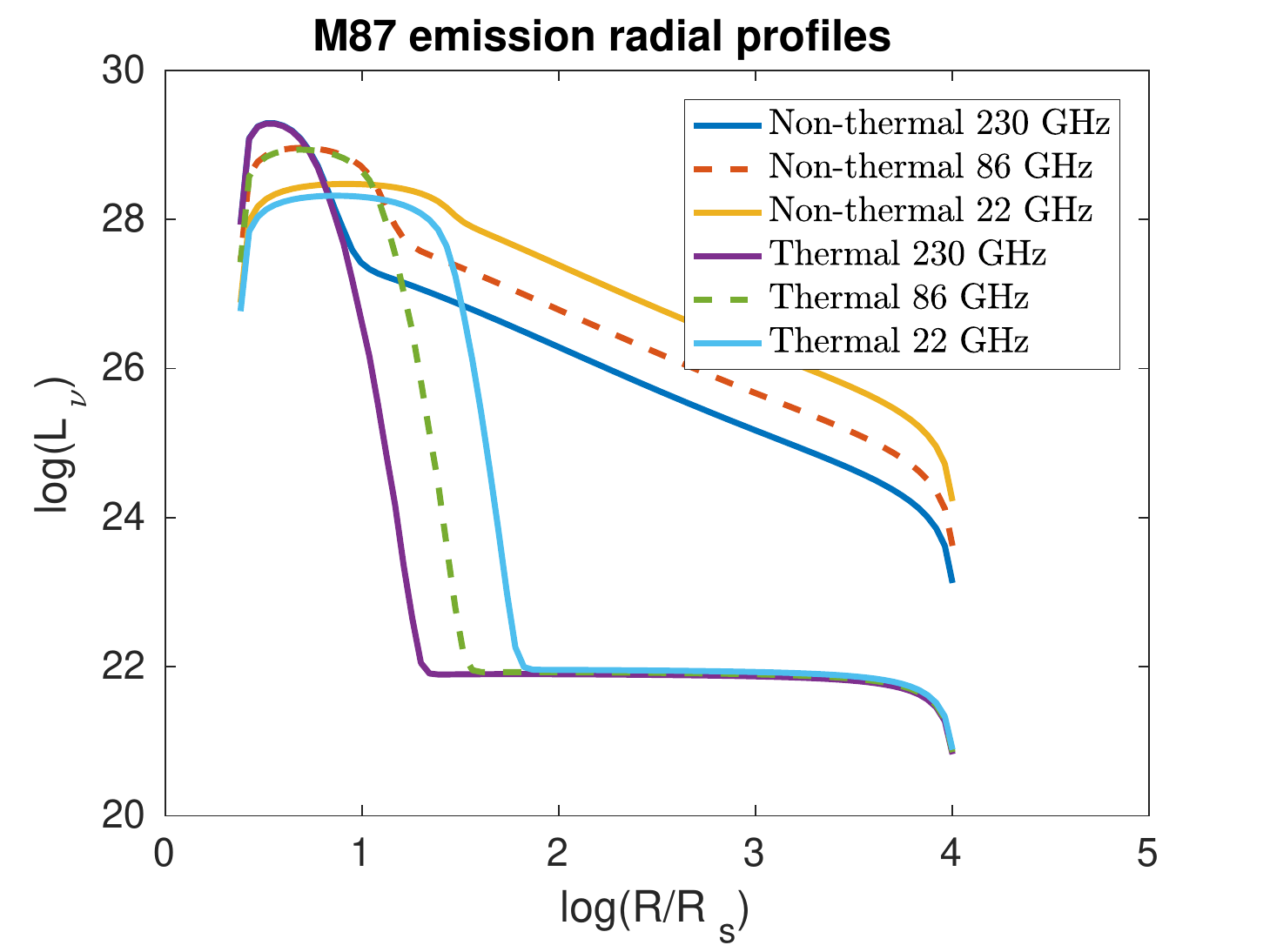}}
 \caption{(a) We show the model SED for M87 for the ADAF, JET and the ADAF+JET models and compare the result with the observational data given in \citet{Prieto16}. (b) The radial profile for emission at 22 GHz, 86 GHz and 230 GHz with the ADAF disk models in cases with and without the inclusion of non-thermal electrons. Here $R_{\rm s}$ is the Schwarzchild radius.} \label{M87spec}
\end{center}
\end{figure*} 
\section{Model Description}
\subsection{Dynamical Equations}
The accretion disks in LLAGN can be described by an advection dominated accretion flow (ADAF). These are sub-Eddington accretion flows \citep{Shapiro76, Ichimaru77, Rees82}, for which the accretion rate is much smaller than the Eddington rate and the gas reaches its virial temperature. Such disks are geometrically thick but optically thin and are often accompanied by outflows. The small accretion rate leads to a lower density in the disk. Such a disk is thus optically thin and the excess heat generated due to viscous dragging is unable to escape due to inefficient radiative cooling, hence it is advected onto the black hole. As a consequence of low opacity, a two-temperature plasma forms, where the ions are much hotter than the electrons. We investigate the evolution of the dynamical equations in an ADAF model tailored to LLAGN \citep{Yuan05}. From the laws of conservation of mass, radial momentum, angular momentum and energy, we set up the following dynamical equations \citep{Yuan14}:

\begin{equation}
  \dot{M}(R)=\dot{M}_{\mathrm{R_{out}}}\left(\frac{R}{R_{\mathrm{out}}}\right)^s=4\pi \rho R H |v| \label{mascon}
\end{equation}

\begin{equation}
 v\frac{dv}{dR}-\Omega^2 R = -\Omega_K^2 R - \frac{1}{\rho}\frac{d}{dR}(\rho c_s^2)
\end{equation}

\begin{equation}
 \frac{d\Omega}{dR} = \frac{v\Omega_K (\Omega R^2 - j)}{\alpha R^2 c_s^2}
\end{equation}
 
 \begin{eqnarray}
  \rho v\left(\frac{de_i}{dR}-\frac{p_i}{\rho^2}\frac{d\rho}{dR}\right) &=& (1-\delta)q^+ - q^{ie} \nonumber \\
   \rho v\left(\frac{de_e}{dR}-\frac{p_e}{\rho^2}\frac{d\rho}{dR}\right) &=& \delta q^+ + q^{ie} - q^-  \label{enercon}
\end{eqnarray}

Here the variables have their usual meaning. It should be noted that eq.[\ref{mascon}] takes into account the case of outflows while eq.[\ref{enercon}] is the modified energy conservation equation for two temperature plasmas. Comparing the modeled spectrum with the available SED data allows us to constrain the  important model parameters like the accretion rate $\dot{M}$, the strength of the outflow parameter $s$, the relative magnetic strength $\beta$ (which is embedded in the energy equation) and the electron heating factor $\delta$ \citep{Xie12, Chael18}.

\subsection{The Jet}
The accretion dynamics in reality is more complex due to turbulence, the presence of magnetic fields, hot spots and outflows. \citet{Narayan94, Narayan95, Blandford99} postulate that hot accretion flows should have strong winds followed by the formation of jets. This is supported by observational evidence which suggests that almost all LLAGN are radio-loud \citep{Falcke00, Nagar00, Ho02}.  The jet dynamics is more complicated but it is accepted to arise from a combination of magnetic fields and rotation. The most accepted models are the Blandford-Znajek (BZ) model \citep{Blandford77} which states that the primary source of energy in the jet is the rotational energy of the black hole and the Blandford-Payne (BP) model \citep{Blandford82} suggests that it is due to the rotational energy of the accretion flow. Independent of the origin of the jet, it is often necessary to include a jet in order to explain the observed SED of most LLAGN \citep{Nemmen14, Li16}. We consider a basic jet model \citep{Spada01, Yuan05} here. The radial velocity of accretion near the supermassive black hole is supersonic and hence the bending of the gas into the jet causes a standing shock at the bottom. From the shock jump condition, post shock properties like the temperature and densities are determined. The shock accelerates a fraction of the electrons yielding a power law energy distribution. It is these electrons which contribute to most of the emission in the jet. The emission from the jet depends on the jet-opening angle $\phi$, the Lorentz factor $\Gamma_j$ , the energy densities $\epsilon_e$ and $\epsilon_B$ for the accelerated electrons and the amplified magnetic fields.

\subsection{The Spectral Energy Distribution (SED) and emission profiles}
The temperature, density of electrons and the velocity profiles of the gas are the parameters that we obtain from the solution of the dynamical equations . Assuming the disk is isothermal in the vertical direction, the spectrum of unscattered photons at a given radius is calculated by solving the radiative transfer equation in the vertical direction of the disk based on the two-stream approximation \citep{Rybicki79}. Since the gas close to the black hole is hot, optically thin and magnetized, the processes which significantly contribute to the emission are synchrotron radiation and Bremsstrahlung \citep{Manmoto97}, while the presence of electrons Comptonize \citep{Coppi90} these photons to modify the total SED. Processes such as magnetic reconnection, weak shocks and turbulent dissipation can accelerate a fraction of the thermal electrons to a non-thermal power-law distribution, which also emits via synchrotron emission \citep{Yuan05}. The power-law electrons in the jet lead to an enhanced contribution of the synchrotron emission.

\begin{figure}
\begin{center}
{ \includegraphics[height=2.2in,width=2.2in]{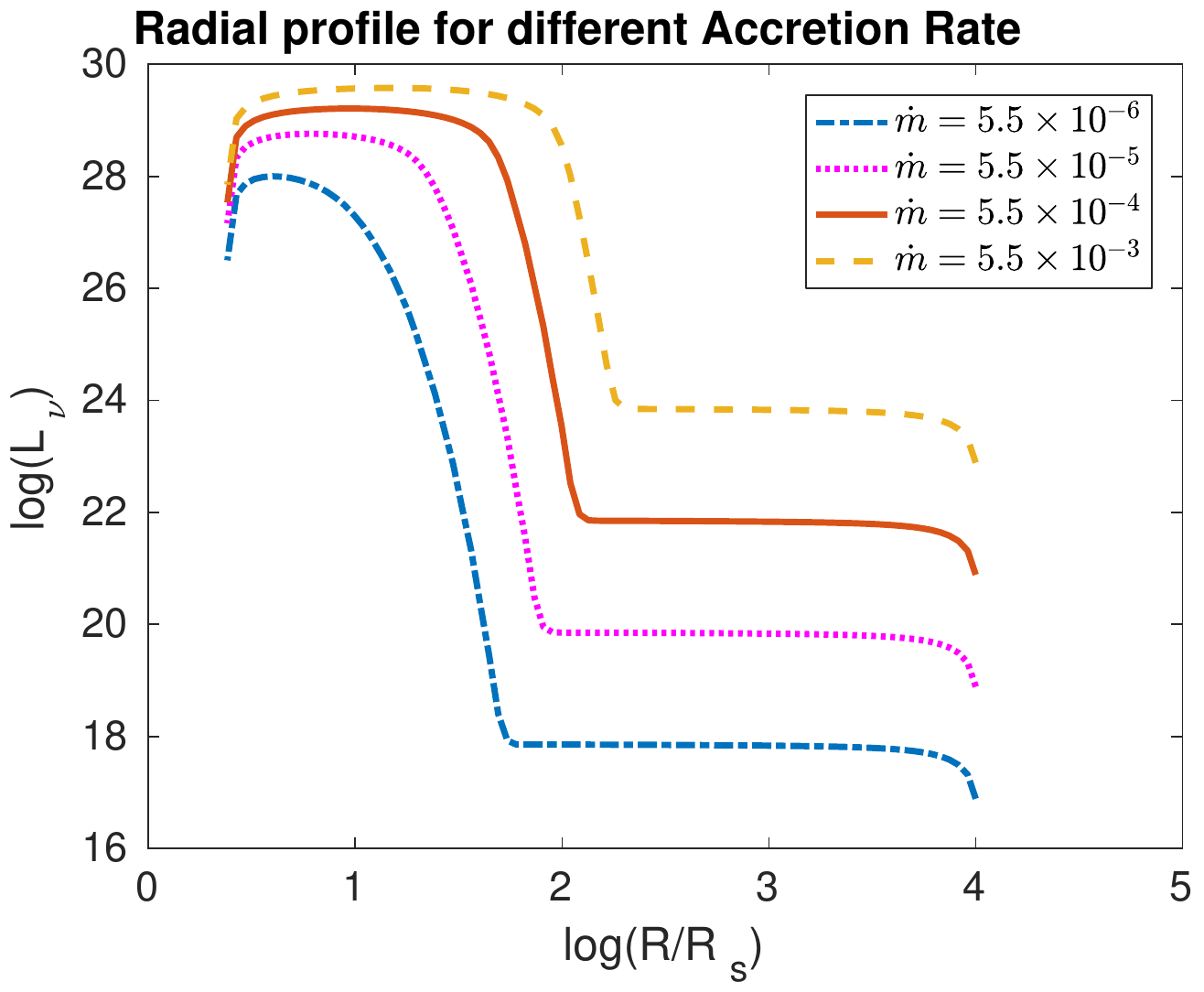}}
 \caption{Variation in the radial profile of emission at $22$ GHz for different accretion rates. Here $\dot{m}=\dot{M}/\dot{M_{\rm Edd} }$ {\it i.e.} the ratio of true accretion rate to the Eddington rate.}\label{accrate} 
\end{center}
\end{figure}
\section{Discussion and Conclusion}
Our primary aim in this work is to find the best fit parameter values of our model by comparing the simulated SED with the observed dataset. As a specific example, we have calculated the SED of M87 and the expected radial profiles of the emission from the disk at three different frequencies (22 GHz, 86 GHz and 230 GHz), which is shown in Fig.~[\ref{M87spec}]. The radial profile changes depending on the parameter values as well as the adopted physics in the disk (mass of the black hole, Eddington ratios, presence of non-thermal electrons, etc.), thereby potentially allowing us to constrain the physics of the accretion disk through a comparison with observations from the EHT (230 GHz) as well as the GMVA (86 GHz) and the EVN (22 GHz). As an example we have shown in Fig.~[\ref{accrate}] the variation in the radial profile of emission from the disk at $22$ GHz for different accretion rates. This analysis is the first step towards predicting the resolvability of the region in the proximity of the black hole. The summary of our analysis for M87 is as follows:
\begin{itemize}
 \item Including the effect of power-law electrons is important especially to explain the Compton peak in the SED.
 \item The radial profile of emission varies significantly if we include the effect of power-law electrons. This shows that the power-law electrons contribute significantly to the emission at low frequencies even from the outer regions of the disk.
 \item For a system like M87 it is important to include the emission from the jet in order to explain the flux at lower frequencies. Any synchrotron emission from the disk will be highly self-absorbed.
\end{itemize}
We plan to apply this analysis to other nearby LLAGN and estimate the detectability of their accretion flows.

\begin{acknowledgement}
We thank for funding via Conicyt, in particular through ALMA-Conicyt (Project No. 31160001), Fondecyt regular (Project No. 1161247), the 'Concurso Proyectos Internacionales de Investigaci\'on, Convocatoria 2015' (project code PII20150171) , Conicyt PIA ACT172033 and the CONICYT project Basal AFB-170002. \end{acknowledgement}
 
 \bibliographystyle{aa}
\small
\bibliography{bidibib}

\begin{thebibliography}{25}
\expandafter\ifx\csname natexlab\endcsname\relax\def\natexlab#1{#1}\fi

\bibitem[{{Blandford} \& {Begelman}(1999)}]{Blandford99}
{Blandford}, R.~D. \& {Begelman}, M.~C. 1999, \mnras, 303, L1

\bibitem[{{Blandford} \& {Payne}(1982)}]{Blandford82}
{Blandford}, R.~D. \& {Payne}, D.~G. 1982, \mnras, 199, 883

\bibitem[{{Blandford} \& {Znajek}(1977)}]{Blandford77}
{Blandford}, R.~D. \& {Znajek}, R.~L. 1977, \mnras, 179, 433

\bibitem[{{Broderick} {et~al.}(2009){Broderick}, {Loeb}, \&
  {Narayan}}]{Broderick09}
{Broderick}, A.~E., {Loeb}, A., \& {Narayan}, R. 2009, ApJ, 701, 1357

\bibitem[{{Broderick} {et~al.}(2015){Broderick}, {Narayan}, {Kormendy},
  {Perlman}, {Rieke}, \& {Doeleman}}]{Broderick15}
{Broderick}, A.~E., {Narayan}, R., {Kormendy}, J., {et~al.} 2015, ApJ, 805, 179

\bibitem[{{Chael} {et~al.}(2018){Chael}, {Rowan}, {Narayan}, {Johnson}, \&
  {Sironi}}]{Chael18}
{Chael}, A., {Rowan}, M., {Narayan}, R., {Johnson}, M., \& {Sironi}, L. 2018,
  \mnras, 478, 5209

\bibitem[{{Coppi} \& {Blandford}(1990)}]{Coppi90}
{Coppi}, P.~S. \& {Blandford}, R.~D. 1990, \mnras, 245, 453

\bibitem[{{Falcke} \& {Markoff}(2000)}]{Falcke00}
{Falcke}, H. \& {Markoff}, S. 2000, \aap, 362, 113

\bibitem[{{Ho}(2002)}]{Ho02}
{Ho}, L.~C. 2002, \apj, 564, 120

\bibitem[{{Ichimaru}(1977)}]{Ichimaru77}
{Ichimaru}, S. 1977, \apj, 214, 840

\bibitem[{{Li} {et~al.}(2016){Li}, {Yuan}, \& {Xie}}]{Li16}
{Li}, Y.-P., {Yuan}, F., \& {Xie}, F.-G. 2016, \apj, 830, 78

\bibitem[{{Manmoto} {et~al.}(1997){Manmoto}, {Mineshige}, \&
  {Kusunose}}]{Manmoto97}
{Manmoto}, T., {Mineshige}, S., \& {Kusunose}, M. 1997, \apj, 489, 791

\bibitem[{{Meyer} {et~al.}(2012){Meyer}, {Ghez}, {Sch{\"o}del}, {Yelda},
  {Boehle}, {Lu}, {Do}, {Morris}, {Becklin}, \&
  {Matthews}}]{2012Sci...338...84M}
{Meyer}, L., {Ghez}, A.~M., {Sch{\"o}del}, R., {et~al.} 2012, Science, 338, 84

\bibitem[{{Nagar} {et~al.}(2000){Nagar}, {Falcke}, {Wilson}, \& {Ho}}]{Nagar00}
{Nagar}, N.~M., {Falcke}, H., {Wilson}, A.~S., \& {Ho}, L.~C. 2000, \apj, 542,
  186

\bibitem[{{Narayan} \& {Yi}(1994)}]{Narayan94}
{Narayan}, R. \& {Yi}, I. 1994, \apjl, 428, L13

\bibitem[{{Narayan} \& {Yi}(1995)}]{Narayan95}
{Narayan}, R. \& {Yi}, I. 1995, \apj, 444, 231

\bibitem[{{Nemmen} {et~al.}(2014){Nemmen}, {Storchi-Bergmann}, \&
  {Eracleous}}]{Nemmen14}
{Nemmen}, R.~S., {Storchi-Bergmann}, T., \& {Eracleous}, M. 2014, \mnras, 438,
  2804

\bibitem[{{Prieto} {et~al.}(2016){Prieto}, {Fern{\'a}ndez-Ontiveros},
  {Markoff}, {Espada}, \& {Gonz{\'a}lez-Mart{\'{\i}}n}}]{Prieto16}
{Prieto}, M.~A., {Fern{\'a}ndez-Ontiveros}, J.~A., {Markoff}, S., {Espada}, D.,
  \& {Gonz{\'a}lez-Mart{\'{\i}}n}, O. 2016, \mnras, 457, 3801

\bibitem[{{Rees} {et~al.}(1982){Rees}, {Begelman}, {Blandford}, \&
  {Phinney}}]{Rees82}
{Rees}, M.~J., {Begelman}, M.~C., {Blandford}, R.~D., \& {Phinney}, E.~S. 1982,
  \nat, 295, 17

\bibitem[{{Rybicki} \& {Lightman}(1979)}]{Rybicki79}
{Rybicki}, G.~B. \& {Lightman}, A.~P. 1979, {Radiative processes in
  astrophysics}

\bibitem[{{Shapiro} {et~al.}(1976){Shapiro}, {Lightman}, \&
  {Eardley}}]{Shapiro76}
{Shapiro}, S.~L., {Lightman}, A.~P., \& {Eardley}, D.~M. 1976, \apj, 204, 187

\bibitem[{{Spada} {et~al.}(2001){Spada}, {Ghisellini}, {Lazzati}, \&
  {Celotti}}]{Spada01}
{Spada}, M., {Ghisellini}, G., {Lazzati}, D., \& {Celotti}, A. 2001, \mnras,
  325, 1559

\bibitem[{{Xie} \& {Yuan}(2012)}]{Xie12}
{Xie}, F.-G. \& {Yuan}, F. 2012, \mnras, 427, 1580

\bibitem[{{Yuan} {et~al.}(2005){Yuan}, {Cui}, \& {Narayan}}]{Yuan05}
{Yuan}, F., {Cui}, W., \& {Narayan}, R. 2005, \apj, 620, 905

\bibitem[{{Yuan} \& {Narayan}(2014)}]{Yuan14}
{Yuan}, F. \& {Narayan}, R. 2014, \araa, 52, 529

\end{thebibliography}


\end{document}